\documentclass[aps,twocolumn,showpacs,preprintnumbers,amsmath,amssymb]{revtex4}
\usepackage{graphicx}% Include figure files
\usepackage{bm}% bold math

\def\br{ \bm{r} }
\def\bR{ \bm{R} }
\def\bk{ \bm{k} }
\def\bK{ \bm{K} }

\def\bq{ \bm{q} }
\def\bA{ \bm{A} }
\def\bB{ \bm{B} }
\def\bv{ \bm{v} }
\def\brho{ \bm{\rho} }

\begin{document}

\title{Magnetic properties of superconductors with strong spin-orbit coupling}

\author{K.~V.~Samokhin}

\affiliation{Department of Physics, Brock University,
St.Catharines, Ontario, Canada L2S 3A1}
\date{\today}

\begin{abstract}
We study the response of a superconductor with a strong spin-orbit
coupling on an external magnetic field. The Ginzburg-Landau free
energy functional is derived microscopically for a general crystal
structure, both with and without an inversion center, and for an
arbitrary symmetry of the superconducting order parameter. As a
by-product, we obtain the general expressions for the intrinsic
magnetic moment of the Cooper pairs. It is shown that the
Ginzburg-Landau gradient energy in a superconductor lacking
inversion symmetry has unusual structure. The general formalism is
illustrated using as an example CePt$_3$Si, which is the first
known heavy-fermion superconductor without an inversion center.
\end{abstract}


\maketitle

\section{Introduction}
\label{sec: Intro}

Superconductors with unconventional, or anisotropic, pairing have
remained one of the favourite and most-studied systems in
condensed matter physics for more than two decades. Any
superconducting material in which the symmetry of the pair wave
function is different from an $s$-wave spin singlet, predicted by
the Bardeen-Cooper-Schrieffer (BCS) theory, can be called
``unconventional''. From the initial discoveries of
superconductivity in the heavy-fermion compounds, the list of
examples has grown to include the high-$T_c$ cuprate
superconductors, ruthenates, magnetic superconductors, and
possibly organic materials. In contrast, such popular novel
superconductor as MgB$_2$, in which the order parameter is an
$s$-wave singlet, is still ``conventional'' despite its many
unusual properties.

Although the pairing mechanism in most if not all unconventional
superconductors is subject to much debate, their behavior can be
well understood using the symmetry approach, pioneered in Refs.
\cite{VG85,UR85,Blount85}. In particular, the intrinsic anisotropy
and the multi-component nature of the order parameter lead to a
variety of interesting magnetic properties, such as the internal
magnetism of the Cooper pairs, multiple phases in the vortex
state, and the upper critical field anisotropy near $T_c$ not
described by the effective mass tensor in the Ginzburg-Landau (GL)
equations, for a review see, e.g., Refs. \cite{Book,LZh95}.

In most of the previous microscopic calculations of the magnetic
properties of unconventional superconductors, the model of an
isotropic band in a centrosymmetric crystal has been used.
Historically, this has its origin in the fact that an
unconventional Cooper pairing was first extensively studied in the
context of the superfluid ${}^3$He, which is indeed an isotropic
Fermi liquid with a weak spin-orbit (SO) coupling \cite{VW90}.
Although taking into account a realistic Fermi surface anisotropy
in a crystalline superconductor is not believed to cause any
drastic qualitative effects, it might lead to some considerable
quantitative changes compared to the parabolic band model. The SO
coupling in crystals is usually taken care of by re-defining the
basis of the single-electron states: instead of the usual Bloch
spinors, the Cooper pairs are now formed by pseudospin eigenstates
\cite{UR85}. Then the only significant change in the
superconducting properties, compared to the case without SO
coupling, is that the system is no longer invariant with respect
to arbitrary $SU(2)$ spin rotations, which alters the symmetry of
the order parameter in the pseudospin-triplet channel
\cite{VG85,UR85}. A detailed analysis of the temperature
dependence of the upper critical field, including the band
anisotropy, impurity scattering, and sometimes the Fermi liquid
corrections, has been done using the quasiclassical (Eilenberger)
method, see e.g. Ref. \cite{CS93} and the references therein. A
disadvantage of this approach is that it assumes a constant
density of states of the normal electrons near the Fermi surface
and therefore fails to capture some contributions to the intrinsic
magnetism of the Cooper pairs. Additional complications arise when
a superconductor with SO interaction lacks an inversion center. In
a nutshell, the symmetry analysis of superconducting phases should
be modified if the SO coupling is strong, because the twofold
degeneracy of the single-electron bands is now lifted almost
everywhere in the Brillouin zone, which makes it impossible to
introduce pseudospin and also suppresses most of the pairing
channels \cite{And84}.

Although most superconductors do have inversion symmetry, there
are some exceptions. Early examples included such materials as
V$_3$Si \cite{AB65} and HfV$_2$ \cite{LZ72}, in which a possible
loss of inversion symmetry is associated with a structural phase
transition in the bulk of the crystal. The existence of
superconductivity was later reported in ferroelectric perovskites
SrTiO$_3$ \cite{SHC64} and BaPbO$_3$-BaBiO$_3$ \cite{BV83}. It was
pointed out in Ref. \cite{GR01} that the surface superconductivity
observed, e.g. in Na-doped WO$_3$ \cite{WO3}, is generically
non-centrosymmetric simply because of the fact that the two sides
of the surface layer are manifestly non-equivalent. Possible
effects of the absence of inversion symmetry in the layered
high-$T_c$ cuprates were discussed in Refs. \cite{Edelstein}. Very
recently, superconductivity was found in non-centrosymmetric
compounds CePt$_3$Si \cite{exp-CePtSi} and UIr \cite{exp-UIr}.

This article is aimed to study the magnetic properties of a clean
superconductor with arbitrary pairing symmetry and band structure,
with or without an inversion center. We focus on the strong SO
coupling limit, which is believed to be the case in many
unconventional superconductors, in particular the heavy-fermion
compounds, because of the presence of elements with large atomic
weights, such as U, Ce, {\em etc}. In contrast to the previous
works, the starting point of our calculations is an effective band
Hamiltonian, which describes the dynamics of the Bloch electrons
in a magnetic field \cite{LL9}. The superconducting pairing is
introduced using a BCS-type weak-coupling model, generalized for
the case of an unconventional pairing symmetry. We derive the GL
free energy microscopically, which allows us not only to calculate
the upper critical field, but also evaluate the intrinsic magnetic
moment of the Cooper pairs in a crystalline superconductor. To the
best of the author's knowledge, a microscopic derivation of the GL
equations for a superconductor lacking an inversion center, in the
presence of an arbitrary SO coupling, has never been done before,
so we fill this gap here. On the other hand, although some of our
results in the centrosymmetric case are not new and can be found
scattered in the literature, we found it instructive to treat both
cases within the same general framework, which also highlights the
important differences between them.

The article is organized as follows. In Section \ref{sec: sp Ham},
we discuss the properties of the Bloch electrons in a magnetic
field in the normal state, and introduce the single-band effective
Hamiltonian. In Section \ref{sec: SC}, we study the properties of
a strong SO coupling superconductor in a magnetic field near
$T_c$, derive the linearized GL equations in the lowest order in
$\bB$, and calculate the internal magnetic moment of the Cooper
pairs, in both the centrosymmetric and non-centrosymmetric cases.
In Section \ref{sec: CePtSi}, we apply the general formalism to
CePt$_3$Si. Section \ref{sec: Conclusion} concludes with a
discussion of our results.

\section{Single-particle properties}
\label{sec: sp Ham}

To develop the necessary framework for the analysis of the
superconducting properties, we first need to understand how a
uniform magnetic field affects the single-electron states in a
normal crystal with SO coupling (with or without an inversion
center). While for free electrons with a parabolic dispersion
$\bm{p}^2/2m$ the magnetic Hamiltonian is obtained by simply
replacing $\bm{p}$ with a gauge-invariant momentum operator
$\bm{p}+(e/c)\bA$ ($e$ is the absolute value of the electron
charge), the case of band electrons should be treated more
carefully.

In zero field, the single-electron Hamiltonian has the form
\begin{equation}
\label{H0 zero B}
    H_0=\sum\limits_{\bk\nu}\epsilon_\nu(\bk)c^\dagger_{\bk\nu}c_{\bk\nu},
\end{equation}
where $c^\dagger$ and $c$ are the creation and annihilation
operators of band electrons with the wave vector $\bk$,
$\epsilon_\nu(\bk)$ is the quasiparticle dispersion in the $\nu$th
band, which takes into account all effects of the periodic lattice
potential and the SO interaction, and $\sum_{\bk}$ stands for the
integration over the first Brillouin zone. We assume that there is
no disorder in the crystal, so that $\bk$ is a good quantum number
in the absence of external fields. The Matsubara Green's function
of electrons, defined in the standard fashion:
\begin{equation}
\label{GF def}
    G_{\nu_1\nu_2}(\bk_1,\tau_1;\bk_2,\tau_2)=
    -\langle T_\tau
    c_{\bk_1\nu_1}(\tau_1)c^\dagger_{\bk_2\nu_2}(\tau_2)\rangle,
\end{equation}
is diagonal with respect to both the band index and the wave
vector:
\begin{equation}
\label{G0}
    G_\nu(\bk,\omega_n)=\frac{1}{i\omega_n-\epsilon_\nu(\bk)},
\end{equation}
where $\omega_n=(2n+1)\pi T$ is the fermionic Matsubara frequency
(we use the units in which $k_B=1$).

In the presence of a nonzero uniform magnetic field $\bB$, Eq.
(\ref{H0 zero B}) is replaced by
\begin{equation}
\label{H0 B}
    H_0=\sum\limits_{\bk\nu}c^\dagger_{\bk\nu}
    {\cal E}_{\nu}(\bk,\bB)c_{\bk\nu},
\end{equation}
where ${\cal E}$ is the effective one-band Hamiltonian in the
$\bk$-space \cite{LL9}. The main technical difficulty in the
derivation of Eq. (\ref{H0 B}) is that the corresponding vector
potential $\bm{A}$ grows linearly as a function of $\br$, leading
to divergent matrix elements of the Hamiltonian with respect to
the zero-field Bloch waves. As was first pointed out by Peierls
\cite{Pei33}, these non-perturbative features can be taken into
account by simply replacing the wave vector $\bk$ in the
zero-field band dispersion $\epsilon_\nu(\bk)$ by the
gauge-invariant combination $\bk+(e/\hbar c)\bm{A}(\hat\br)$,
where $\hat\br=i\bm{\nabla}_{\bk}$ is the position operator in the
$\bk$-representation. Later, this idea was elaborated in Refs.
\cite{LKW,BlountRoth}, where it was shown that the Peierls
Hamiltonian corresponds in fact to the zero-order term in the
expansion of the general effective one-band Hamiltonian in powers
of $\bB$:
\begin{equation}
\label{Heff}
    {\cal E}_{\nu}(\bk,\bB)=\epsilon_\nu(\bK)+
    B_i \epsilon_{\nu,i}^{(1)}(\bK)+
    B_i B_j \epsilon_{\nu,ij}^{(2)}(\bK)+...,
\end{equation}
where $\bK$ is an operator in the $\bk$-space:
$$
    \bK=\bk+\frac{e}{\hbar
    c}\bm{A}(\hat\br)=
    \bk+i\frac{e}{2\hbar c}\left(\bB\times\frac{\partial}{\partial\bk}\right)
$$
[here and below we use the symmetric gauge:
$\bA=(\bB\times\br)/2$]. Since the components of $\bK$ do not
commute: $[K_i,K_j]=-i(e/\hbar c)e_{ijk}B_k$, the order of
application is important, so that ${\cal E}$ is assumed to be a
completely symmetrized function of $K_i$. This can be achieved,
e.g., by representing the expansion coefficients in Eq.
(\ref{Heff}), which are periodic in $\bk$, in the form of a
Fourier series over the lattice vectors $\bm{R}$, and then
replacing $\bk\to\bK$ to obtain the operators
$\epsilon_\nu(\bK)=\sum_{\bm{R}}\tilde\epsilon_\nu(\bm{R})e^{-i\bm{R}\bK}$,
{\em etc}.

If the electron bands are degenerate in zero field due to spin or
pseudospin (see Sec. \ref{sec: sp Ham I} below), then the
effective Hamiltonian ${\cal E}$ and all the expansion
coefficients are $2\times 2$ matrices. The Green's function
corresponding to the Hamiltonian (\ref{H0 B}) is not diagonal with
respect to $\bk$, because the system is no longer invariant under
lattice translations (it is still invariant though under the
magnetic translations which combine the lattice translations with
gauge transformations).

Although the explicit expressions for the expansion coefficients
in Eq. (\ref{Heff}) can be derived, at least in principle, using
the procedure described in detail in Refs. \cite{BlountRoth}, some
important information can be obtained from general symmetry
considerations. The full symmetry group ${\cal G}$ of the system
in the normal state is given by a product of the space group and
the gauge group $U(1)$. Assuming that there is no magnetic order
in zero field and omitting the lattice translations, we can write
${\cal G}=G\times K\times U(1)$, where $G$ is the point group of
the crystal, which may or may not include the inversion operation
$I$, and $K$ is the time reversal operation. At non-zero $\bB$,
the Hamiltonian (\ref{H0 B}) is invariant with respect to time
reversal only if the sign of $\bB$ (and of $\bA$) is also changed,
which imposes the following constraint on the function ${\cal E}$:
$K^\dagger{\cal E}_{\nu}(-\bB)K={\cal E}_{\nu}(\bB)$. In addition,
the expansion coefficients must have certain transformation
properties under the action of the point group elements, in
particular, the band dispersion $\epsilon(\bk)$ must be invariant
under all operations from $G$.

Further steps depend crucially on whether or not there is an
inversion center in the crystal lattice, which determines the
degeneracy of the zero-field bands.

\subsection{Crystals with inversion center}
\label{sec: sp Ham I}

If the crystal has an inversion center, then the bands are
two-fold degenerate at each $\bk$, because the Bloch states
$\psi_{\bk +}=\psi_{\bk\nu}$ and $\psi_{\bk -}=KI\psi_{\bk\nu}$
have the same energy, belong to the same wave vector, and are
orthogonal. These states can be chosen to transform under the
action of the space group operations similar to the spin
eigenstates, in which case they are referred to as the pseudospin
states \cite{UR85}. Thus the bands can be labelled by
$\nu=(n,\alpha)$, where $\alpha=\pm$ is the pseudospin projection.
Focussing on a single band, we can omit the index $n$, and the
effective band Hamiltonian (\ref{Heff}) becomes
\begin{equation}
\label{Heff 1}
    {\cal
    E}_{\alpha\beta}(\bk,\bB)=\epsilon(\bK)\delta_{\alpha\beta}-
    B_i\mu_{ij}(\bK)\sigma_{j,\alpha\beta}+...,
\end{equation}
where $\sigma_j$ are the Pauli matrices, and both the zero-field
band dispersion $\epsilon(\bk)$ and the tensor $\mu_{ij}(\bk)$ are
invariant under all point group operations. It is easy to see that
this form of the effective Hamiltonian is compatible with all the
symmetry requirements, in particular that ${\cal E}$ should be
Hermitian and $K$- and $I$-invariant. Indeed, the time reversal
operator is $K=(i\sigma_2)K_0$, where $K_0$ is the operation of
complex conjugation, which changes $\bk\to -\bk$. Therefore, we
have $[\sigma_2{\cal E}(-\bk,-\bB)\sigma_2]^*={\cal E}(\bk,\bB)$.
Also, ${\cal E}(-\bk,\bB)={\cal E}(\bk,\bB)$, because of inversion
symmetry. In the limit of zero SO coupling, the usual Zeeman
interaction term is recovered: $\mu_{ij}(\bk)\to
\mu_B\delta_{ij}$, where $\mu_B$ is the Bohr magneton.

The Green's function (\ref{GF def}) is a $2\times 2$ matrix in the
pseudospin space, which satisfies the following equation in the
frequency representation:
\begin{equation}
\label{GFequation k}
 (i\omega_n-{\cal E}_1)_{\alpha\gamma}G_{\gamma\beta}(\bk_1,\bk_2;\omega_n)
 =\delta_{\alpha\beta}\delta(\bk_1-\bk_2).
\end{equation}
The Fourier transform of the Green's function, defined as
\begin{equation}
\label{G rr def}
    G_{\alpha\beta}(\br_1,\br_2;\omega_n)=\sum\limits_{\bk_1,\bk_2}
    e^{i\bk_1\br_1-i\bk_2\br_2}
    G_{\alpha\beta}(\bk_1,\bk_2;\omega_n),
\end{equation}
satisfies the equation
\begin{equation}
\label{GFequation r}
 (i\omega_n-\hat{\cal E}_1)_{\alpha\gamma}
 G_{\gamma\beta}(\br_1,\br_2;\omega_n)=\delta_{\alpha\beta}\delta(\br_1-\br_2),
\end{equation}
where $\hat{\cal E}$ is the Fourier transform of the effective
band Hamiltonian (\ref{Heff 1}), which is obtained by simply
replacing $\bK$ by the real space operator
$$
    \hat\bK=-i\frac{\partial}{\partial\br}+\frac{e}{\hbar
    c}\bm{A}(\br)=-i\frac{\partial}{\partial\br}+\frac{e}{2\hbar
    c}(\bB\times\br).
$$
The subscript 1 in ${\cal E}_1$ or $\hat{\cal E}_1$ means that the
operator acts on the first argument of $G$. It should be noted
that the Green's function (\ref{G rr def}) is not the same as the
Green's function of the band electrons in the coordinate
representation. The latter is defined as
$\langle\br_1\sigma|(i\omega_n-H_0)^{-1}|\br_2\sigma'\rangle=
\langle\br_1\sigma|\bk_1\alpha\rangle
G_{\alpha\beta}(\bk_1,\bk_2;\omega_n)\langle\bk_2\beta|\br_2\sigma'\rangle$
(the summation over repeated indices is implied), where
$\langle\br\sigma|\bk\alpha\rangle=\psi_{\bk\alpha}(\br\sigma)$ is
the Bloch spinor, with $\sigma=\uparrow,\downarrow$ being the
$z$-projection of spin.

The second term in $\hat\bK$ presents some difficulty because it
grows linearly as a function of $\br$ and therefore cannot be
treated as a small perturbation. To handle this problem, we seek
solution of Eq. (\ref{GFequation r}) in a factorized form
\begin{equation}
\label{GF factorized 1}
 G_{\alpha\beta}(\br_1,\br_2;\omega_n)=
 \bar G_{\alpha\beta}(\br_1-\br_2,\omega_n)e^{i\varphi(\br_1,\br_2)}
\end{equation}
where $\varphi(\br_1,\br_2)=(e/\hbar
c)\int_{\br_1}^{\br_2}\bm{A}(\br)d\br$, and the integration goes
along a straight line connecting $\br_1$ and $\br_2$ \cite{Gor59}.
Using the identities
\begin{equation}
\label{dvarphidr}
 \frac{\partial}{\partial\br_{1,2}}\int_{\br_1}^{\br_2}\bm{A}(\br)d\br
 =\mp\bm{A}(\br_{1,2})+\frac{1}{2}[\bB\times(\br_1-\br_2)],
\end{equation}
one can show that the translationally-invariant function
$\bar G(\br_1-\br_2)=\bar G(\bR)$ obeys the equation
\begin{equation}
\label{barG eq R}
    (i\omega_n-\bar{\cal E})_{\alpha\gamma}\bar
    G_{\gamma\beta}(\bR,\omega_n)=\delta_{\alpha\beta}\delta(\bR),
\end{equation}
where the operator $\bar{\cal E}$ is obtained by replacing
$\hat\bK$ in the argument of $\hat{\cal E}$ in Eq.
(\ref{GFequation r}) by
$\hat\bK_{\bR}=-i\partial/\partial\bR+(e/2\hbar c)(\bB\times\bR)$.

The advantage of introducing the function $\bar G$ is that, in
contrast to Eq. (\ref{GFequation r}), the magnetic field term in
Eq. (\ref{barG eq R}) can be treated as a perturbation at small
enough $\bB$. The precise condition can be easily obtained in the
case of an isotropic parabolic band
$\epsilon(\bk)=\hbar^2(k^2-k_F^2)/2m$, when the solution of Eq.
(\ref{barG eq R}) in zero field is $\bar
G_{\alpha\beta}(\bR,\omega_n)\sim\delta_{\alpha\beta}
e^{ik_FR\,\mathrm{sign}\,\omega_n}e^{-|\omega_n|R/v_F}$, where
$v_F=\hbar k_F/m$ is the Fermi velocity. Because of the fast
oscillations of the exponential, the characteristic scale of the
derivative $\partial/\partial\bR$ is $k_F$. On the other hand, the
scale of $R$ is given by $\hbar v_F/k_BT$, so that the
field-dependent term in $\hat\bK_{\bR}\bar G$ is small compared
with the gradient term if $\hbar\omega_c\ll k_BT$, where
$\omega_c=eB/mc$ is the cyclotron frequency. Although this
condition does not have a simple form for a realistic band
structure, it is usually assumed that the perturbative treatment
of $\bB$ in Eq. (\ref{barG eq R}) is legitimate at all but very
low temperatures, where the Landau level quantization effects
become important.

The Fourier transform of $\bar G$ satisfies the equation
\begin{equation}
\label{barG eq k}
 \left[i\omega_n-\bar{\cal E}(\bk,\bB)\right]_{\alpha\gamma}
 \bar G_{\gamma\beta}(\bk,\omega_n)=\delta_{\alpha\beta},
\end{equation}
which is solved perturbatively in $\bB$. The expansion of the
effective band Hamiltonian has the form
\begin{equation}
\label{bar cal E expansion}
    \bar{\cal E}_{\alpha\beta}(\bk,\bB)=\epsilon(\bk)\delta_{\alpha\beta}
    -\bB\bm{m}_{\alpha\beta}(\bk)+O(B^2),
\end{equation}
where
\begin{equation}
\label{m def}
    m_{i,\alpha\beta}(\bk)=i\frac{e}{2c}
    \left[\bv(\bk)\times\frac{\partial}{\partial\bk}\right]_i\delta_{\alpha\beta}+
    \mu_{ij}(\bk)\sigma_{j,\alpha\beta}.
\end{equation}
The first term comes from the expansion of $\epsilon(\bK)$, with
$\bv(\bk)=(1/\hbar)\partial\epsilon(\bk)/\partial\bk$ being the
band velocity, while the second one is obtained by replacing $\bK$
with $\bk$. As obvious from Eq. (\ref{bar cal E expansion}),
$\bm{m}$ can be interpreted as the magnetic moment operator of the
band electrons, although one cannot say that the first and the
second terms correspond to the orbital and the spin magnetic
moments respectively, because both $\bv(\bk)$ and $\mu_{ij}(\bk)$
include the effects of SO coupling. The solution of Eq. (\ref{barG
eq k}) can be written as $\bar G=\bar G_0-\bB\bar G_0\bm{m}\bar
G_0+O(B^2)$. Inserting the expression (\ref{m def}) here and
keeping only the corrections of the first order in $\bB$, we have
\begin{equation}
\label{barG expansion}
    \bar G_{\alpha\beta}(\bk,\omega_n)=
    \frac{\delta_{\alpha\beta}}{i\omega_n-\epsilon(\bk)}-B_i\mu_{ij}(\bk)
    \frac{\sigma_{j,\alpha\beta}}{[i\omega_n-\epsilon(\bk)]^2}.
\end{equation}
Note that because of inversion symmetry, $\bar
G_{\alpha\beta}(-\bk,\omega_n)=\bar
G_{\alpha\beta}(\bk,\omega_n)$.

\subsection{Crystals without inversion center}

In the absence of inversion center in the crystal lattice, the
electron bands are non-degenerate almost everywhere, except from
some high-symmetry lines in the Brillouin zone. The formal reason
for this is that without the inversion operation $I$, one cannot
in general construct two orthogonal degenerate Bloch states at the
same $\bk$ (note that the Kramers theorem still holds: there is a
degeneracy between the time reversed states $\psi_{\bk\nu}$ and
$K\psi_{\bk\nu}$ belonging to $\bk$ and $-\bk$ respectively). The
above is not valid at zero SO coupling. In that case, there is an
additional symmetry in the system -- the invariance with respect
to arbitrary spin rotations, which leads to the bands being
two-fold degenerate because of spin, so that the results of the
previous section apply.

Assuming that the SO coupling is strong and the bands are well
split (which is the case in CePt$_3$Si \cite{SZB04}), the
effective single-band Hamiltonian (\ref{Heff}) can be written in
the following form
\begin{equation}
\label{Heff 2}
    {\cal E}(\bk,\bB)=\epsilon(\bK)-\bB\bm{\lambda}(\bK)+...,
\end{equation}
where the band dispersion $\epsilon(\bk)$ is invariant with
respect to all point group operations, and $\bm{\lambda}(\bk)$ is
a pseudovector, which, being a property of the crystal in zero
field, satisfies the conditions
$(g\bm{\lambda})(g^{-1}\bk)=\bm{\lambda}(\bk)$, where $g$ is any
operation from the point group. Because of the time-reversal
symmetry, we also have $\epsilon(-\bk)=\epsilon(\bk)$ and
$\bm{\lambda}(-\bk)=-\bm{\lambda}(\bk)$. At a nonzero $\bB$ we
have ${\cal E}(-\bk,-\bB)={\cal E}(\bk,\bB)$, but ${\cal
E}(-\bk,\bB)\neq{\cal E}(\bk,\bB)$ in general, because of the lack
of inversion symmetry. An example of the microscopic calculation
of $\bm{\lambda}(\bk)$ using a simple two-dimensional model is
given at the end of this subsection. Also, in Section \ref{sec:
CePtSi} below, we discuss how to find the momentum dependence of
$\bm{\lambda}$ in a non-centrosymmetric tetragonal crystal.

The only modification to the analysis of Sec. \ref{sec: sp Ham I}
is that both the effective Hamiltonian (\ref{Heff}) and the
Green's function (\ref{GF def}) become scalar functions. The
Green's function is factorized:
\begin{equation}
\label{GF factorized 2}
    G(\br_1,\br_2;\omega_n)=\bar G(\br_1-\br_2,\omega_n)
    e^{i\varphi(\br_1,\br_2)},
\end{equation}
where the Fourier transform of $\bar G$ satisfies the equation
\begin{equation}
\label{barG eq 2}
 \left[i\omega_n-\bar{\cal E}(\bk,\bB)\right]\bar G(\bk,\omega_n)=1.
\end{equation}
As in the centrosymmetric case, at low fields we solve this
equation perturbatively in $\bB$, using
\begin{equation}
\label{cal E expansion}
    \bar{\cal E}(\bk,\bB)=\epsilon(\bk)
    -\bB\bm{m}(\bk)+O(B^2),
\end{equation}
where
\begin{equation}
\label{lambda k def 2}
    \bm{m}(\bk)=i\frac{e}{2c}
    \left[\bv(\bk)\times\frac{\partial}{\partial\bk}\right]+\bm{\lambda}(\bk)
\end{equation}
has the meaning of the magnetic moment operator of the band
electrons. The contribution from the first term in $\bm{m}$ to
$\bar G$ vanishes, and we finally have, in the first order in
$\bB$,
\begin{equation}
\label{barG expansion 2}
    \bar G(\bk,\omega_n)=
    \frac{1}{i\omega_n-\epsilon(\bk)}
    -B_i \lambda_i(\bk)\frac{1}{[i\omega_n-\epsilon(\bk)]^2}.
\end{equation}

Most of the previous works on non-centrosymmetric superconductors,
both two-dimensional \cite{Edelstein,GR01,DF03}, and
three-dimensional \cite{FAKS04}, have been based on the Rashba
model (we would like to mention, in particular, Ref.
\cite{Edel96}, in which the GL functional was derived for a
one-component $s$-wave order parameter in a Rashba
superconductor). In this model, the combined effect of the SO
coupling and the lack of inversion symmetry is mimicked by an
additional term in the single-particle Hamiltonian:
\begin{equation}
\label{Rashba}
    H_0=\sum\limits_{\bk}\epsilon_0(\bk)a^\dagger_{\bk\sigma}a_{\bk\sigma}
    +\gamma\sum\limits_{\bk}\bm{n}\cdot(\bm{\sigma}_{\sigma\sigma'}
    \times\bk)\,a^\dagger_{\bk\sigma}a_{\bk\sigma'}.
\end{equation}
Here $\sigma,\sigma'=\uparrow,\downarrow$ is the $z$-axis spin
projection, the operator $a_{\bk\sigma}$ destroys an electron in a
Bloch state of energy $\epsilon_0(\bk)$ corresponding to zero SO
coupling, and $\bm{n}$ is a unit vector allowed by symmetry (in a
2D system, $\bm{n}$ is simply the normal vector to the plane).
Choosing $\bm{n}=\hat z$, we diagonalize the Hamiltonian
(\ref{Rashba}) by a unitary transformation
$a_{\bk\sigma}=U_{\bk,\sigma n}c_{\bk n}$ ($n=1,2$), which gives
two Rashba bands:
\begin{equation}
\label{Rashba bands}
    \epsilon_{1(2)}(\bk)=\epsilon_0(\bk)\pm|\gamma|k_\perp
\end{equation}
($k_\perp=\sqrt{k_x^2+k_y^2}$), with the eigenfunctions
\begin{equation}
\label{Rashba eigenfunctions}
    \psi_{\bk,1(2)}(\br)=\frac{1}{\sqrt{2}}\left(%
    \begin{array}{c}
    1 \\
    \mp ie^{i\varphi_{\bk}} \\
    \end{array}%
    \right)e^{i\bk\br},
\end{equation}
where $\tan\varphi_{\bk}=k_y/k_x$. The bands (\ref{Rashba bands})
are non-degenerate almost everywhere, touching only at the two
poles of the Fermi surface along the $z$ axis. We would like to
emphasize that the band indices $n=1,2$ cannot be interpreted as
the pseudospin projections. Indeed, under time reversal the
pseudospin eigenstates would transform similar to the spin
eigenstates, i.e. into one another. However, being a symmetry of
the Hamiltonian time reversal transforms the Rashba bands into
themselves, which can be directly verified for the eigenstates
(\ref{Rashba eigenfunctions}):
\begin{eqnarray*}
    K\psi_{\bk,1}=(i\sigma_2)K_0\psi_{\bk,1}=
    \frac{ie^{-i\varphi_{\bk}}}{\sqrt{2}}\left(%
    \begin{array}{c}
    1 \\
    ie^{i\varphi_{\bk}} \\
    \end{array}%
    \right)e^{-i\bk\br}\\
    =\frac{ie^{-i\varphi_{\bk}}}{\sqrt{2}}\left(%
    \begin{array}{c}
    1 \\
    -ie^{i\varphi_{-\bk}} \\
    \end{array}%
    \right)e^{-i\bk\br}\propto\psi_{-\bk,1},
\end{eqnarray*}
and similarly for $\psi_{\bk,2}$ (we used
$\varphi_{-\bk}=\varphi_{\bk}+\pi$).

It is easy to show that in the presence of a non-zero magnetic
field the effective Hamiltonian for the Rashba model can be cast
in the form (\ref{Heff 2}). To obtain the pseudovector
$\bm{\lambda}(\bk)$, let us consider a two-dimensional system in a
field parallel to the $xy$ plane. Then the Hamiltonian
(\ref{Rashba}) is modified by the Zeeman term:
$H_B=H_0-\mu_B\bm{\sigma}\bB$. The diagonalization of $H_B$,
followed by an expansion in powers of $B$, gives
\begin{eqnarray*}
    {\cal E}_{1(2)}(\bk,\bB)&=&\epsilon_0(\bk)\\
    &&\pm\sqrt{\gamma^2k_\perp^2+2\gamma\mu_B(\bk\times\bB)_x
    +\mu_B^2B^2}\\
    &\simeq& \epsilon_{1(2)}(\bk)-\bm{\lambda}_{1(2)}(\bk)\bB,
\end{eqnarray*}
where
\begin{equation}
\label{lambda Rashba}
    \bm{\lambda}_{1(2)}(\bk)=\pm\mu_B\frac{\bk\times\bm{n}}{k_\perp}.
\end{equation}
In this article, we want to keep our discussion as general as
possible and therefore do not resort to any explicit model, such
as the Rashba model, to describe the SO coupling. Our results are
based only on the symmetry considerations and valid for an
arbitrary strength of the SO coupling and any band structure.

\section{Magnetic response in the superconducting state}
\label{sec: SC}

\subsection{Crystals with inversion center}
\label{sec: GL I}

Now let us take into account the attractive interaction between
the band electrons in the Cooper channel. The total Hamiltonian is
given by $H=H_0+H_{int}$, where the free electron Hamiltonian
$H_0$ is given by Eq. (\ref{H0 B}) and, for a BCS-type mechanism
of pairing, the interaction part can be written as
\begin{eqnarray}
\label{H BCS 1}
    H_{int}=\frac{1}{2}\sum\limits_{\bk,\bk',\bq}
    V_{\alpha\beta,\gamma\delta}(\bk,\bk')c^\dagger_{\bk+\bq/2,\alpha}
    c^\dagger_{-\bk+\bq/2,\beta}\nonumber\\
    \times c_{-\bk'+\bq/2,\gamma}c_{\bk'+\bq/2,\delta}.
\end{eqnarray}
The pairing potential does not depend on the external magnetic
field and is assumed to have a factorized form:
\begin{equation}
\label{pairing potential 1}
 V_{\alpha\beta,\gamma\delta}(\bk,\bk')=-\frac{1}{2}V
 \sum\limits_{a=1}^{d}\Psi_{a,\alpha\beta}(\bk)
 \Psi^\dagger_{a,\gamma\delta}(\bk').
\end{equation}
with the coupling constant $V>0$. Here $\Psi_a(\bk)$ are the
$2\times 2$ matrix basis functions of an irreducible
representation $\Gamma$ of dimensionality $d$ of the symmetry
group of the system at zero magnetic field \cite{Book}. The
pairing interaction is nonzero only inside a thin shell of width
$\epsilon_c$ (the cutoff energy) in the vicinity of the Fermi
surface $\epsilon(\bk)=0$, i.e.
$\Psi_a(\bk)=\Psi_a(\bk_F)f_c[\epsilon(\bk)]$, where $\bk_F$ is a
wave vector at the Fermi surface and the cutoff function
$f_c(\epsilon)$ is localized about the origin, e.g.
$f_c(\epsilon)=\theta(\epsilon_c-|\epsilon|)$. The basis functions
are assumed to be orthonormal:
\begin{eqnarray}
\label{normal cond}
    &&\frac{1}{2}\left\langle\mathrm{tr}\,
    [\Psi_a^\dagger(\bk)\Psi_b(\bk)]\right\rangle_\epsilon\nonumber\\
    &&\qquad=\frac{1}{2}\left\langle\mathrm{tr}\,
    [\Psi_a^\dagger(\bk)\Psi_b(\bk)]\right\rangle_0f_c^2(\epsilon)
    =\delta_{ab}f_c^2(\epsilon),\qquad
\end{eqnarray}
where the angular brackets denote the averaging over the constant
energy surface $\epsilon(\bk)=\epsilon$:
\begin{equation}
\label{FS average}
    \langle(...)\rangle_\epsilon=\frac{1}{N_0(\epsilon)}
    \sum\limits_{\bk}(...)\delta[\epsilon-\epsilon(\bk)],
\end{equation}
and $N_0(\epsilon)=\sum_{\bk}\delta[\epsilon-\epsilon(\bk)]$ is
the normal-metal density of states (DoS) per one pseudospin
projection.

It follows from anti-commutation of the fermionic operators that
$\Psi_{a,\beta\alpha}(-\bk)=-\Psi_{a,\alpha\beta}(\bk)$. In the
presence of inversion symmetry, the even in $\bk$
(pseudospin-singlet) and odd in $\bk$ (pseudospin-triplet) pairing
states can be considered separately. In the singlet case, the
matrix basis functions can be represented in the form
\begin{equation}
\label{singlet OP}
    \Psi_{a,\alpha\beta}(\bk)=(i\sigma_2)_{\alpha\beta}\phi_a(\bk),
\end{equation}
where $\phi_a(\bk)$ are the even scalar basis functions of the
$\Gamma$ representation. In the triplet case, we have
\begin{equation}
\label{triplet OP}
    \Psi_{a,\alpha\beta}(\bk)=(i\sigma_i\sigma_2)_{\alpha\beta}\phi_{a,i}(\bk)
\end{equation}
where $\bm{\phi}_a(\bk)$ are the odd vector basis functions of the
$\Gamma$ representation \cite{VG85,Book}.

The superconducting order parameter can be represented as a linear
combination of the basis functions:
\begin{equation}
\label{OP definition 1}
    \Delta_{\alpha\beta}(\bk,\bq)=\sum\limits_a\eta_a(\bq)\Psi_{a,\alpha\beta}(\bk),
\end{equation}
with the coefficients $\eta_a$ playing the role of the order
parameter components, which determine, for instance, the free
energy ${\cal F}$ of the superconductor. In the vicinity of the
critical temperature $T_c(\bB)$, one can keep only the quadratic
terms in the expansion of ${\cal F}$:
\begin{equation}
\label{F r}
    {\cal F}=\sum\limits_{ab}\int d\br\;\eta_a^*(\br)\,S_{ab}\,\eta_b(\br).
\end{equation}
Here $S$ is a $d\times d$ matrix differential operator of infinite
order:
\begin{equation}
\label{S}
    S_{ab}=\frac{1}{V}\delta_{ab}-
    \int d\bR\;\bar S_{ab}(\bR)e^{-i\bR\bm{D}},
\end{equation}
where $\bm{D}=-i\nabla_{\br}+(2e/\hbar c)\bm{A}$, and the
translationally-invariant function $\bar S_{ab}(\bR)$ is expressed
in terms of the Green's functions (\ref{barG eq R}). Its Fourier
transform is given by
\begin{widetext}
\begin{eqnarray}
\label{bar S}
    \bar S_{ab}(\bq)&=&T\sum\limits_n\sum\limits_{\bk}
    \Lambda^{ab}_{\alpha\beta\gamma\delta}(\bk)
    \bar G_{\beta\gamma}\left(\bk+\frac{\bq}{2},\omega_n\right)
    \bar
    G_{\alpha\delta}\left(-\bk+\frac{\bq}{2},-\omega_n\right),
\end{eqnarray}
\begin{eqnarray*}
    \Lambda^{ab}_{\alpha\beta\gamma\delta}(\bk)&=&\frac{1}{2}
    \exp\left[i\frac{e}{4\hbar c}\bB\left(
    \frac{\partial}{\partial\bk_1}\times\frac{\partial}{\partial\bk_2}
    \right)\right]
    \Psi_{a,\alpha\beta}^\dagger(\bk_1)\Psi_{b,\gamma\delta}(\bk_2)
    \biggr|_{\bk_1=\bk_2=\bk}\nonumber\\
    &=&\frac{1}{2}\Psi_{a,\alpha\beta}^\dagger(\bk)
    \Psi_{b,\gamma\delta}(\bk)+i\frac{e}{8\hbar c}\bB
    (\nabla_{\bk}\Psi_{a,\alpha\beta}^\dagger\times
    \nabla_{\bk}\Psi_{b,\gamma\delta})+O(B^2).
\end{eqnarray*}
\end{widetext}
The derivation of these formulas is outlined in Appendix \ref{app:
Proof F r}. As obvious from Eq. (\ref{S}), the operator $S$ is a
completely symmetrized function of the components of $\bm{D}$,
which do not commute: $[D_i,D_j]=-i(2e/\hbar c)e_{ijk}B_k$. Also,
its Taylor expansion contains only even powers of $\bm{D}$,
because $\bar S_{ab}(-\bR)=\bar S_{ab}(\bR)$ due to the inversion
symmetry.

The field dependence of the phase transition temperature at
arbitrary $\bB$ can be found from Eq. (\ref{F r}): $T_c(\bB)$ is
defined as the temperature at which the minimum eigenvalue of the
operator $S$ passes through zero. For an isotropic $s$-wave order
parameter, the corresponding equations were derived and solved in
Refs. \cite{Hc2 refs}, while for an isotropic $p$-wave case it was
done in Ref. \cite{SK80}. In a general case, i.e. for an arbitrary
band structure and pairing symmetry, $T_c(\bB)$ can only be
calculated numerically.

Here we focus on the properties of our superconductor in the weak
field limit. We have ${\cal F}=\int F\,d\br$, where the free
energy density can be represented as
\begin{equation}
\label{F density}
    F=A_{ab}\eta_a^*\eta_a+K_{ab,ij}\eta_a^*D_iD_j\eta_b-\bm{M}\bB.
\end{equation}
This expression has the usual form expected on the
phenomenological grounds, with $K_{ab,ij}$ being the generalized
effective mass tensor, and $\bm{M}$ having the meaning of the
intrinsic magnetic moment of the Cooper pairs. The linearized GL
equations follow from Eq. (\ref{F density}) after the minimization
over the order parameter: $\delta{\cal F}/\delta\eta_a^*(\br)=0$.
Below we outline how to calculate the free energy density using
our weak-coupling model.

The first term in $F$ is obtained by putting $\bq=\bB=0$ in Eqs.
(\ref{S}) and (\ref{bar S}), which gives
\begin{equation}
\label{f0 B0}
    A_{ab}=\frac{1}{V}\delta_{ab}
    -\frac{1}{2}\sum\limits_{\bk}
    \mathrm{tr}\,[\Psi_a^\dagger(\bk)\Psi_b(\bk)]{\cal S}[\epsilon(\bk)],
\end{equation}
and
$$
    {\cal S}(\epsilon)=T\sum\limits_n\frac{1}{\omega_n^2+\epsilon^2}=
    \frac{1}{2\epsilon}\tanh\frac{\epsilon}{2T}.
$$
The necessary momentum cutoff in Eq. (\ref{f0 B0}) is provided by
the basis functions $\Psi_a(\bk)$, which are restricted to the
vicinity of the Fermi surface. Calculating the momentum integral
with the help of the normalization condition (\ref{normal cond}),
we obtain $A_{ab}=[(1/V)-I]\delta_{ab}$, where
\begin{equation}
\label{I}
    I(T)=\int d\epsilon\;
    N_0(\epsilon)f^2_c(\epsilon){\cal S}(\epsilon)\simeq
    N_F\ln\frac{2e^C\epsilon_c}{\pi T}
\end{equation}
($C\simeq 0.577$ is Euler's constant). To obtain this result we
made the usual assumption that $N_0(\epsilon)$ is a slowly-varying
function within the energy shell of width $\epsilon_c$ near the
Fermi surface, which allows us to replace it by a constant -- the
DoS at the Fermi level $N_F=N_0(0)$. At the zero-field critical
temperature $T_c$, we have $I(T_c)=1/V$, which gives the standard
BCS result: $T_c\simeq 1.13\epsilon_c\exp(-1/N_FV)$. Expanding
$A_{ab}$ in the vicinity of $T_c$, we recover the familiar
expression for the uniform term in the free energy density:
\begin{equation}
\label{F 0}
    A_{ab}=\alpha(T-T_c)\delta_{ab},
\end{equation}
where $\alpha=N_F/T_c$.

Next, we calculate the intrinsic magnetic moment $\bm{M}$. Using
the small-$\bB$ expansions of the normal-state Green's function
$\bar G$ and the vertex $\Lambda$, we obtain in the singlet case:
\begin{equation*}
    M_i=\frac{ie}{4\hbar c}\eta^*_a\eta_b\left\langle\left(
    \nabla_{\bk}\phi^*_a\times\nabla_{\bk}\phi_b\right)_i\right\rangle_0
    I,
\end{equation*}
where $\langle(...)\rangle_0$ stands for the Fermi-surface
averaging (\ref{FS average}), and $I$ is defined by Eq. (\ref{I}).
To derive this expression, we again used the fact that the basis
functions are non-zero only in a narrow vicinity of the Fermi
surface, which allows one to separate the energy integration from
the integration over the Fermi surface. A similar calculation in
the triplet case gives
\begin{eqnarray*}
    M_i&=&\frac{ie}{4\hbar c}\eta^*_a\eta_b\left\langle\left(
    \nabla_{\bk}\bm{\phi}^*_{a}\times\nabla_{\bk}\bm{\phi}_{b}\right)_i
    \right\rangle_0 I\nonumber\\
    &&+2i\eta^*_a\eta_b\left\langle\mu_{ij}(\bk)
    \left(\bm{\phi}_a^*\times\bm{\phi}_b\right)_j\right\rangle_0
    I_1,
\end{eqnarray*}
where
\begin{eqnarray}
\label{I 1}
    I_1(T)=\int d\epsilon\;
    N_0(\epsilon)f^2_c(\epsilon){\cal S}_1(\epsilon)
    \simeq-\frac{N'_F}{2}\ln\frac{2e^C\epsilon_c}{\pi T},\\
    \nonumber {\cal S}_1(\epsilon)=T\sum\limits_n
    \frac{1}{(i\omega_n-\epsilon)^2}\frac{1}{-i\omega_n-\epsilon}
    =\frac{1}{2}\frac{\partial S(\epsilon)}{\partial\epsilon}.
\end{eqnarray}
Here $N'_F=N'_0(0)$ is a measure of the electron-hole asymmetry
near the Fermi surface. Putting $T=T_c$, using the BCS result for
the critical temperature, and choosing real basis functions (which
can always be done if the normal state is non-magnetic) we finally
obtain the density of the intrinsic magnetic moment of the Cooper
pairs:
\begin{equation}
\label{M final}
    \bm{M}=i\bm{\gamma}_{ab}\eta^*_a\eta_b,
\end{equation}
where $\bm{\gamma}_{ab}=-\bm{\gamma}_{ba}$ is given by
\begin{equation}
\label{gamma singlet}
    \gamma_{i,ab}=\frac{e}{4\hbar c}\frac{1}{V}e_{ijl}
    \left\langle\frac{\partial\phi_a(\bk)}{\partial k_j}
    \frac{\partial\phi_b(\bk)}{\partial k_l}\right\rangle_0
\end{equation}
in the singlet case, and
\begin{eqnarray}
\label{gamma triplet}
    \gamma_{i,ab}=\frac{e}{4\hbar c}\frac{1}{V}
    e_{ijl}\left\langle\frac{\partial\phi_{a,m}(\bk)}{\partial k_j}
    \frac{\partial\phi_{b,m}(\bk)}{\partial k_l}
    \right\rangle_0\nonumber\\
    -\frac{N'_F}{N_F}\frac{1}{V}e_{jkl}\left\langle\mu_{ij}(\bk)
    \phi_{a,k}(\bk)\phi_{b,l}(\bk)\right\rangle_0
\end{eqnarray}
in the triplet case. It follows from these expressions that
$\bm{M}=0$ for any order parameter corresponding to a
one-dimensional representation of the point group, both in the
singlet and triplet cases.

Finally, let us evaluate the gradient terms in Eq. (\ref{F
density}). The magnetic field dependence of the coefficients
$K_{ab,ij}$ can be neglected, which follows from the fact that the
lowest eigenvalue of the operator $K_{ab,ij}D_iD_j$ is already
linear in $|\bB|$, see Appendix \ref{app: Linear B}. The physical
meaning of this is simple: the suppression of the critical
temperature due to the gradient energy is always linear in a weak
field, regardless of the dimensionality of the order parameter and
the shape of the Fermi surface. Taking the second order derivative
in Eq. (\ref{bar S}) at $\bB=0$ and calculating the Matsubara
sums, we obtain:
\begin{eqnarray*}
    K_{ab,ij}=-\frac{1}{4}\hbar^2\left\langle\mathrm{tr}\,
    [\Psi_a^\dagger(\bk)\Psi_b(\bk)]
    v_i(\bk)v_j(\bk)\right\rangle_0I_2\nonumber\\
    -\frac{1}{8}\hbar^2\left\langle\mathrm{tr}\,
    [\Psi_a^\dagger(\bk)\Psi_b(\bk)]m^{-1}_{ij}(\bk)
    \right\rangle_0I_1.
\end{eqnarray*}
Here $m^{-1}_{ij}(\bk)=(1/\hbar^2)\partial^2\epsilon(\bk)/\partial
k_j\partial k_j$ is the inverse tensor of effective masses, $I_1$
is defined by Eq. (\ref{I 1}), and
\begin{equation}
\label{I 2}
    I_2(T)=\int d\epsilon\; N_0(\epsilon)f^2_c(\epsilon)
    {\cal S}_2(\epsilon)\simeq -\frac{7\zeta(3)}{8\pi^2T^2}N_F,\\
\end{equation}
where
\begin{eqnarray*}
    {\cal S}_2(\epsilon)&=&T\sum\limits_n\biggl[
    \frac{1}{(i\omega_n-\epsilon)^3}\frac{1}{-i\omega_n-\epsilon}\\
    &&-\frac{1}{2}\frac{1}{(i\omega_n-\epsilon)^2}\frac{1}{(-i\omega_n-\epsilon)^2}
    \biggr]\\
    &=&-\frac{1}{16T^2\epsilon}\sinh\left(\frac{\epsilon}{2T}\right)
    \cosh^{-3}\left(\frac{\epsilon}{2T}\right).
\end{eqnarray*}
has a peak near $\epsilon=0$, and $\zeta(s)$ is Riemann's
zeta-function.

Putting all the pieces together, replacing $T$ with $T_c$, and
using real basis functions, we finally have
\begin{eqnarray}
\label{Kabij singlet}
    K_{ab,ij}=\frac{7\zeta(3)\hbar^2}{16\pi^2T_c^2}N_F\left\langle\phi_a(\bk)\phi_b(\bk)
    v_i(\bk)v_j(\bk)\right\rangle_0\nonumber\\
    +\frac{\hbar^2}{8V}\frac{N'_F}{N_F}\left\langle\phi_a(\bk)\phi_b(\bk)m^{-1}_{ij}(\bk)
    \right\rangle_0
\end{eqnarray}
in the singlet case, and
\begin{eqnarray}
\label{Kabij triplet}
    K_{ab,ij}=\frac{7\zeta(3)\hbar^2}{16\pi^2T_c^2}N_F\left\langle\phi_{a,l}(\bk)
    \phi_{b,l}(\bk)v_i(\bk)v_j(\bk)\right\rangle_0\nonumber\\
    +\frac{\hbar^2}{8V}\frac{N'_F}{N_F}\left\langle\phi_{a,l}(\bk)\phi_{b,l}(\bk)
    m^{-1}_{ij}(\bk)\right\rangle_0
\end{eqnarray}
in the triplet case. Assuming a spherical Fermi surface, a
completely isotropic pairing corresponding to the unity
representation of $G$, and neglecting the electron-hole asymmetry,
Eq. (\ref{Kabij singlet}) yields
$K_{ij}=\delta_{ij}[7\zeta(3)\hbar^2/48\pi^2T_c^2]N_Fv_F^2$
\cite{Gor59}. For an anisotropic Fermi surface, but still a
conventional pairing, the results of Ref. \cite{GMB63} are
recovered.

Now we would like to make a few comments about our results. The
internal magnetism of superconductors has been discussed before
mostly for a charged isotropic Fermi liquid without SO coupling,
see, e.g. Ref. \cite{BM85}. In this case, the density of the pair
magnetic moment can be divided into the orbital and spin parts,
both being small due to the smallness of both the quasiclassical
parameter $(k_F\xi_0)^2\ll 1$ ($\xi_0$ is the coherence length),
and the electron-hole asymmetry $N_F'$ \cite{Book}. Here we do not
make any assumptions about the strength of the SO coupling,
therefore the orbital and the spin magnetic moments cannot be
separated, in general. For a general band dispersion, one can
neglect neither of these contributions {\em apriori}, before
calculating the Fermi-surface averages in Eqs. (\ref{gamma
singlet}) and (\ref{gamma triplet}). In particular, the energy
dependence of the single-electron DoS in the metals with $d$- and
$f$-electrons is usually quite significant, which can lead to an
appreciable electron-hole asymmetry near the Fermi level.

In terms of the response of the superconductor on a weak external
field, the gradient terms produce a linear in $\bB$ suppression of
$T_c$, see Appendix \ref{app: Linear B}. The value of the slope
$dH_{c2}/dT$ can be calculated either analytically (in very few
cases), or using a variational approach. On the other hand, the
pair magnetism can compete with the gradient energy, leading even
to the possibility of increasing $T_c$ as a function of $\bB$, if
the internal magnetic moment is large enough. Such mechanism was
recently proposed in Ref. \cite{WS02} to explain the phase diagram
of the ferromagnetic superconductor ZrZn$_2$.

\subsection{Crystals without inversion center}
\label{sec: GL no I}

In this case, the calculations are somewhat simpler because the
bands are non-degenerate. We assume that the Cooper pairing occurs
only between the electrons in the states with opposite momenta,
which are transformed into each other by time reversal. Then the
most general BCS-type Hamiltonian can be written in the form
\begin{equation}
\label{H123}
    H_{int}=H^{(1)}_{int}+H^{(2)}_{int}+H^{(3)}_{int},
\end{equation}
where
\begin{eqnarray*}
    &&H^{(1)}_{int}=\frac{1}{2}\sum\limits_{n}\sum\limits_{\bk,\bk'}
    V^{(1)}_n(\bk,\bk')c^\dagger_{\bk n}c^\dagger_{-\bk n}
    c_{-\bk' n}c_{\bk' n}\\
    &&H^{(2)}_{int}=\frac{1}{2}\sum\limits_{n\neq m}\sum\limits_{\bk,\bk'}
    V^{(2)}_{nm}(\bk,\bk')c^\dagger_{\bk n}c^\dagger_{-\bk n}
    c_{-\bk' m}c_{\bk' m}\\
    &&H^{(3)}_{int}=\frac{1}{2}\sum\limits_{n\neq m}\sum\limits_{\bk,\bk'}
    V^{(3)}_{nm}(\bk,\bk')c^\dagger_{\bk n}c^\dagger_{-\bk m}
    c_{-\bk' m}c_{\bk' n}.
\end{eqnarray*}
Here $n$ and $m$ label the non-degenerate single-electron bands,
e.g. the Rashba bands (\ref{Rashba bands}). The Hamiltonian
$H^{(1)}_{int}$ describes the intra-band pairing, $H^{(2)}_{int}$
describes the pair scattering between the bands, which can result
in the superconducting gaps induced on more than one sheet of the
Fermi surface, and $H^{(3)}_{int}$ corresponds to the pairing of
electrons from different bands.

A considerable simplification occurs if the superconducting gaps
are much smaller than the interband energies. For example, the
band structure calculations of Ref. \cite{SZB04} show that the SO
band splitting in CePt$_3$Si exceeds the superconducting gap by
orders of magnitude. In this situation, the formation of interband
pairs described by $H^{(3)}_{int}$ is strongly suppressed for the
same reasons as for the paramagnetically limited singlet
superconductors \cite{CC62}: the interband splitting cuts off the
logarithmic singularity in the Cooper channel, thus reducing the
critical temperature. Although the bands may touch at some
isolated points at the Fermi surface, as is the case for the
Rashba bands (\ref{Rashba bands}) at $\bk\parallel\hat z$, the
interband pairing in the vicinity of those points is still
suppressed due to the phase space limitations. We also neglect the
possibility of the Cooper pairs having a non-zero momentum
(Larkin-Ovchinnikov-Fulde-Ferrell phase) \cite{LOFF}, which is
expected to be suppressed as well by the large depairing effect of
the SO band splitting.

In this paper, we further neglect the inter-band pair scattering
process described by $H^{(2)}_{int}$, leaving the investigation of
its effects for future work. Thus, we focus on a single
non-degenerate band for which the pairing between time-reversed
states $|\bk\rangle$ and $K|\bk\rangle\sim|-\bk\rangle$ near the
Fermi surface can be written as
\begin{equation}
\label{H BCS 2 0}
    H_{int}=\frac{1}{2}\sum\limits_{\bk,\bk'}
    \tilde
    V(\bk,\bk')c^\dagger_{\bk}c^\dagger_{K\bk}c_{K\bk'}c_{\bk'},
\end{equation}
where $c^\dagger_{K\bk}$ denotes the creation operator of an
electron in the state $K|\bk\rangle$, and the pairing potential is
assumed to have a factorized form
\begin{equation}
\label{pairing potential 2}
 \tilde V(\bk,\bk')=-V
 \sum\limits_{a=1}^{d}\phi_a(\bk)
 \phi^*_a(\bk').
\end{equation}
with $V>0$. Here $\phi_a(\bk)$ are the scalar basis functions of
an irreducible representation $\Gamma$ of the point group of the
crystal in the absence of magnetic field, which are nonzero only
inside the energy shell of width $\epsilon_c$ near the Fermi
surface: $\phi_a(\bk)=\phi_a(\bk_F)f_c[\epsilon(\bk)]$, and
orthonormal:
\begin{equation}
\label{normal cond 2}
   \left\langle
   \phi_a^*(\bk)\phi_b(\bk)\right\rangle_\epsilon=\left\langle
   \phi_a^*(\bk)\phi_b(\bk)\right\rangle_0f_c^2(\epsilon)=
   \delta_{ab}f_c^2(\epsilon).
\end{equation}
The parity of the basis functions can be determined using the
following arguments \cite{SC04}. Although the time-reversed state
$K|\bk\rangle$ belongs to the wave vector $-\bk$, it is not the
same as $|-\bk\rangle$. In fact,
$K|\bk\rangle=t(\bk)|-\bk\rangle$, where $t(\bk)$ is a non-trivial
phase factor, which satisfies $t(-\bk)=-t(\bk)$. This allows us to
write $c^\dagger_{K\bk}=t(\bk)c^\dagger_{-\bk}$ and
$c_{K\bk}=t^*(\bk)c_{-\bk}$. Inserting these relations in Eq.
(\ref{H BCS 2 0}), we have
\begin{equation}
\label{H BCS 2 1}
    H_{int}=\frac{1}{2}\sum\limits_{\bk,\bk'}
    V(\bk,\bk')c^\dagger_{\bk}c^\dagger_{-\bk}c_{-\bk'}c_{\bk'},
\end{equation}
where $V(\bk,\bk')=t(\bk)t^*(\bk')\tilde V(\bk,\bk')$. From the
anti-commutation of fermionic operators it follows that $\tilde
V(\bk,\bk')$ has to be an even function of both arguments, i.e.
one should choose even basis functions $\phi_a(\bk)$ in the
expansion (\ref{pairing potential 2}). Treating the interaction
(\ref{H BCS 2 1}) in the mean-field approximation, one obtains the
order parameter $\Delta(\bk)=t(\bk)\sum_a\eta_a\phi_a(\bk)$, which
is odd in $\bk$. In Ref. \cite{SZB04}, the nodal structure of
$\Delta(\bk)$ was analyzed in terms of the odd basis functions.
This has been corrected in Ref. \cite{SC04}, where the the
importance of the phase factor $t(\bk)$ was recognized.

Allowing for the possibility of a non-uniform superconducting
order parameter, the Hamiltonian (\ref{H BCS 2 1}) becomes
\begin{eqnarray}
\label{H BCS 2}
    H_{int}=\frac{1}{2}\sum\limits_{\bk,\bk',\bq}
    V(\bk,\bk')c^\dagger_{\bk+\bq/2}c^\dagger_{-\bk+\bq/2}\nonumber\\
    \times c_{-\bk'+\bq/2}c_{\bk'+\bq/2}.
\end{eqnarray}
The order parameter can be represented as
\begin{equation}
\label{OP definition 2}
    \Delta(\bk,\bq)=\sum\limits_a\eta_a(\bq)\Psi_a(\bk),
\end{equation}
where $\Psi_a(\bk)=t(\bk)\phi_a(\bk)=-\Psi_a(-\bk)$ satisfy the
orthonormality condition
$\langle\Psi_a^*(\bk)\Psi_b(\bk)\rangle_\epsilon=\delta_{ab}f_c^2(\epsilon)$,
see Eq. (\ref{normal cond 2}).

The contribution to the free energy quadratic in the order
parameter has the form (\ref{F r}) with the kernel now given by
\begin{equation}
\label{S 2}
    S_{ab}=\frac{1}{2V}\delta_{ab}-\frac{1}{2}
    \int d\bR\;\bar S_{ab}(\bR)e^{-i\bR\bm{D}},
\end{equation}
where $S_{ab}(\bR)$ is the Fourier transform of
\begin{widetext}
\begin{eqnarray}
\label{bar S 2}
    \bar S_{ab}(\bq)&=&T\sum\limits_n\sum\limits_{\bk}
    \Lambda_{ab}(\bk)
    \bar G\left(\bk+\frac{\bq}{2},\omega_n\right)
    \bar G\left(-\bk+\frac{\bq}{2},-\omega_n\right),
\end{eqnarray}
\begin{eqnarray*}
    \Lambda_{ab}(\bk)&=&
    \exp\left[i\frac{e}{4\hbar c}\bB\left(
    \frac{\partial}{\partial\bk_1}\times\frac{\partial}{\partial\bk_2}\right)\right]
    \Psi_{a}^*(\bk_1)\Psi_{b}(\bk_2)\biggr|_{\bk_1=\bk_2=\bk}\nonumber\\
    &=&\Psi_{a}^*(\bk)
    \Psi_{b}(\bk)+i\frac{e}{4\hbar c}\bB
    (\nabla_{\bk}\Psi_{a}^*\times
    \nabla_{\bk}\Psi_{b})+O(B^2).
\end{eqnarray*}
\end{widetext}
The derivation is similar to the centrosymmetric case, see
Appendix \ref{app: Proof F r}.

An important difference from the previous case is that, although
the functions $\Psi_a(\bk)$ still have a definite parity, the
Green's functions (\ref{barG expansion 2}) do not: $\bar
G(-\bk,\omega_n)\neq\bar G(\bk,\omega_n)$ in general, therefore
$\bar S_{ab}(-\bR)\neq\bar S_{ab}(\bR)$. This means that the
expansion of the free energy density now contains gradient terms
of an odd degree in $\bm{D}$:
\begin{equation}
\label{F expansion}
    F=
    f^{(0)}_{ab}\,\eta_a^*\eta_b+
    f^{(1)}_{ab,i}\;\eta_a^*D_i\eta_b+f^{(2)}_{ab,ij}\;\eta_a^*D_iD_j\eta_b+... ,
\end{equation}
where
\begin{eqnarray*}
    && f^{(0)}_{ab}=\frac{1}{2V}\delta_{ab}-\bar S_{ab}(\bq=\bm{0}),\\
    && f^{(1)}_{ab,i}=-
    \left.\frac{\partial\bar S_{ab}(\bq)}{\partial
    q_i}\right|_{\bq=\bm{0}},\\
    && f^{(2)}_{ab,ij}=-\frac{1}{2}
    \left.\frac{\partial^2\bar S_{ab}(\bq)}{\partial q_i\partial
    q_j}\right|_{\bq=\bm{0}},\\
    && etc.
\end{eqnarray*}
Using Eq. (\ref{barG expansion 2}), it is easy to see that
$f^{(1)}_{ab,i}=0$ at $\bB=0$.

Keeping only the lowest order terms in the free energy density
expansion in a weak field, we have:
\begin{eqnarray}
\label{F density 2}
    F=A_{ab}\eta_a^*\eta_b+K_{ab,ij}\,\eta_a^*D_iD_j\eta_b-\bm{M}\bB\nonumber\\
    +\tilde K_{ab,ij}B_i\,\eta_a^*D_j\eta_b,
\end{eqnarray}
where $\tilde K_{ab,ij}=\partial f^{(1)}_{ab,j}/\partial
B_i|_{T=T_c,\bB=0}$. The uniform contribution to $F$ can be
calculated in the same fashion as in the previous section, and we
obtain
\begin{equation}
\label{F 0 2}
    A_{ab}=\alpha(T-T_c)\delta_{ab},
\end{equation}
where the critical temperature $T_c$ is given by the same BCS
expression as in the centrosymmetric case, but now
$\alpha=N_F/2T_c$.

The pair magnetic moment $\bm{M}$ and the generalized effective
mass tensor $K_{ab,ij}$ can be calculated similarly to the
centrosymmetric case. Using real basis functions $\phi_a(\bk)$, we
obtain
\begin{equation}
\label{M final 2}
    \bm{M}=i\frac{e}{8\hbar c}\frac{1}{V}e_{ijl}
    \left\langle\frac{\partial\Psi^*_a(\bk)}{\partial k_j}
    \frac{\partial\Psi_b(\bk)}{\partial
    k_l}\right\rangle_0\eta^*_a\eta_b,
\end{equation}
and
\begin{eqnarray}
\label{Kabij 2}
    K_{ab,ij}=\frac{7\zeta(3)\hbar^2}{32\pi^2T_c^2}N_F\left\langle\phi_a(\bk)\phi_b(\bk)
    v_i(\bk)v_j(\bk)\right\rangle_0\nonumber\\
    +\frac{\hbar^2}{16V}\frac{N'_F}{N_F}\left\langle\phi_a(\bk)\phi_b(\bk)m^{-1}_{ij}(\bk)
    \right\rangle_0.
\end{eqnarray}
To calculate the coefficient $\tilde K_{ab,ij}$, we expand $\bar
S_{ab}(\bq)$ to the first order in both $\bB$ and $\bq$ and
evaluate the Matsubara sums, which gives
\begin{eqnarray*}
    \tilde K_{ab,ij}&=&\hbar\left\langle\phi_a^*(\bk)\phi_b(\bk)
    \lambda_i(\bk)v_j(\bk)\right\rangle_0I_2\nonumber\\
    &&+\frac{1}{2}\left\langle\phi_a^*(\bk)\phi_b(\bk)
    \frac{\partial\lambda_i(\bk)}{\partial k_j}
    \right\rangle_0I_1,
\end{eqnarray*}
where $\bm{\lambda}(\bk)$ is the momentum-dependent pseudovector
that determines the linear response of the band electrons on a
weak magnetic field, see Eq. (\ref{Heff 2}), and $I_{1,2}$ are
defined by Eqs. (\ref{I 1}) and (\ref{I 2}) respectively. Using
real basis functions, we finally have
\begin{eqnarray}
\label{tilde Kabij final}
    \tilde K_{ab,ij}=-\frac{7\zeta(3)\hbar}{8\pi^2T_c^2}N_F
    \left\langle\phi_a(\bk)\phi_b(\bk)
    \lambda_i(\bk)v_j(\bk)\right\rangle_0\nonumber\\
    -\frac{1}{4V}\frac{N'_F}{N_F}\left\langle\phi_a(\bk)\phi_b(\bk)
    \frac{\partial\lambda_i(\bk)}{\partial k_j}\right\rangle_0.
\end{eqnarray}
Note that the phase factors $t(\bk)$ have dropped out of both
$K_{ab,ij}$ and $\tilde K_{ab,ij}$. To evaluate the Fermi-surface
averages in Eqs. (\ref{M final 2},\ref{Kabij 2},\ref{tilde Kabij
final}) explicitly, one has to know the band structure [including
$\bm{\lambda}(\bk)$ and $t(\bk)$] and the momentum dependence of
the order parameter.

\section{Applications to C\lowercase{e}P\lowercase{t}$_3$S\lowercase{i}}
\label{sec: CePtSi}

CePt$_3$Si is a heavy-fermion material without inversion center,
which was recently found to become superconducting at $T\simeq
0.75$K \cite{exp-CePtSi}. It has a tetragonal lattice symmetry
described by the point group $G=\mathbf{C}_{4v}$, which is
generated by the rotations $C_{4z}$ about the $z$ axis by an angle
$\pi/2$ and the reflections $\sigma_x$ in the vertical plane
$(100)$. The Fermi surface is invariant under all the operations
from $\mathbf{C}_{4v}$ and also the inversion, the latter being
the consequence of the time-reversal symmetry. The band structure
calculations of Ref. \cite{SZB04} show that the SO coupling in
this material is strong and therefore the degeneracy of the bands
is lifted everywhere, except along the $z$ axis.

\begin{table}
\caption{\label{Table} The character table and the examples of the
basis functions of the irreducible representations of
$\mathbf{C}_{4v}$.}
\begin{ruledtabular}
\begin{tabular}{|c|c|c|c|c|c|}
   $\Gamma$   & $E$ & $C_{4z}$ & $\sigma_x$ & even $\phi_{\Gamma}(\bk)$
                   & odd $\phi_{\Gamma}(\bk)$\\ \hline
   $A_1$      & 1   & 1 & 1 & $k_x^2+k_y^2+ck_z^2$ & $k_z$ \\ \hline
   $A_2$      & 1   & 1 & $-1$ & $(k_x^2-k_y^2)k_xk_y$ & $(k_x^2-k_y^2)k_xk_yk_z$ \\ \hline
   $B_1$      & 1   & $-1$ & 1 & $k_x^2-k_y^2$ & $(k_x^2-k_y^2)k_z$ \\ \hline
   $B_2$      & 1   & $-1$ & $-1$ & $k_xk_y$ & $k_xk_yk_z$ \\ \hline
   $E$        & 2   & 0 & 0 & $k_xk_z$\ ,\ $k_yk_z$ & $k_x$\ ,\ $k_y$
\end{tabular}
\end{ruledtabular}
\end{table}

The point group $\mathbf{C}_{4v}$ has five irreducible
representations: four one-dimensional ($A_1$, $A_2$, $B_1$, and
$B_2$), and one two-dimensional ($E$), see Table \ref{Table}.
Although the order parameter is odd in $\bk$ \cite{SZB04}, its
nodal structure is determined by the even basis functions
\cite{SC04}. Here we consider only the case of a one-component
order parameter, for which
\begin{equation}
\label{OP CPS}
    \Delta(\bk,\br)=\eta(\br)\Psi(\bk)=\eta(\br)t(\bk)\phi(\bk),
\end{equation}
where $\phi(\bk)=\phi(-\bk)$. The pair magnetic moment vanishes,
and the GL free energy (\ref{F density 2}) takes the form
$$
    F=\alpha(T-T_c)|\eta|^2+K_{ij}\eta^*D_iD_j\eta
    +\tilde K_{ij}B_i\eta^*D_j\eta.
$$
Dropping the terms proportional to $N_F'$ and using the symmetry
of the Fermi surface, we have
\begin{equation}
\label{Kij CPS}
\begin{array}{l}
    \displaystyle K_{xx}=K_{yy}=K_1=
    \frac{7\zeta(3)\hbar^2}{32\pi^2T_c^2}N_F\left\langle\phi^2(\bk)
    v_x^2(\bk)\right\rangle_0,\\
    \displaystyle K_{zz}=K_2=\frac{7\zeta(3)\hbar^2}{32\pi^2T_c^2}N_F\left\langle\phi^2(\bk)
    v_z^2(\bk)\right\rangle_0.
\end{array}
\end{equation}

In order to calculate $\tilde K_{ij}$, we need an expression for
$\bm{\lambda}(\bk)$, which satisfies the following symmetry
requirements: $\bm{\lambda}(-\bk)=-\bm{\lambda}(\bk)$,
$(C_{4z}\bm{\lambda})(C_{4z}^{-1}\bk)=\bm{\lambda}(\bk)$, and
$(\sigma_x\bm{\lambda})(\sigma_x^{-1}\bk)=\bm{\lambda}(\bk)$
(since $\bm{\lambda}$ is a pseudovector, we have
$\sigma_x\bm{\lambda}\equiv
IC_{2x}\bm{\lambda}=C_{2x}\bm{\lambda}$, where $C_{2x}$ is a
rotation by an angle $\pi$ about the $x$ axis). To solve these
constraints, we represent $\bm{\lambda}$ as an expansion over the
odd basis functions of the irreducible representations of
$\mathbf{C}_{4v}$, see Table \ref{Table}:
\begin{equation}
\label{lambda CPS expansion}
    \bm{\lambda}(\bk)=\sum\limits_\Gamma\sum\limits_{a=1}^{d_\Gamma}
    \bm{\lambda}_{\Gamma,a}\tilde\phi_{\Gamma,a}(\bk),
\end{equation}
where $\tilde\phi(-\bk)=-\tilde\phi(\bk)$. It is straightforward
to check that only the representations $A_2$ and $E$ contribute to
the expansion (\ref{lambda CPS expansion}), so that the most
general expression for $\bm{\lambda}(\bk)$, which satisfies all
the symmetry requirements, is given by
\begin{equation}
\label{lambda CPS general}
    \bm{\lambda}(\bk)=\lambda_E\left[\tilde\phi_{E,2}(\bk)\hat x-
    \tilde\phi_{E,1}(\bk)\hat y\right]+\lambda_{A_2}\tilde\phi_{A_2}(\bk)\hat
    z,
\end{equation}
where $\lambda_E$ and $\lambda_{A_2}$ are constants. Substituting
it into Eq. (\ref{tilde Kabij final}), using the fact that the
Fermi velocity $\bm{v}(\bk)$ transforms according to a vector
representation $E+A_1$, and dropping the terms proportional to
$N_F'$, we finally have
\begin{eqnarray}
\label{tilde Kij CPS}
    &&\tilde K_{xy}=-\tilde K_{yx}=\tilde K\nonumber\\
    &&\quad =-\frac{7\zeta(3)\hbar}{8\pi^2T_c^2}N_F\left\langle\phi^2(\bk)\tilde\phi_{E,1}(\bk)
    v_x(\bk)\right\rangle_0.
\end{eqnarray}
All other $\tilde K_{ij}$ vanish by symmetry.

Finally, the GL free energy density can be written as
\begin{eqnarray}
\label{GL CPS final}
    F=\alpha(T-T_c)|\eta|^2+\eta^*\left[K_1(D_x^2+D_y^2)+
    K_2D_z^2\right]\eta\nonumber\\
    +\tilde K\eta^*(B_xD_y-B_yD_x)\eta.
\end{eqnarray}
While the second-order gradient terms here are typical for a
one-component order parameter in a uniaxial crystal, the last,
linear in both $\bm{D}$ and $\bB$, term is unusual and occurs only
because of the absence of inversion symmetry.

As an application of the above results, let us calculate the upper
critical fields for $\bB$ parallel and perpendicular to the $z$
axis. To this end, we solve the linearized GL equation obtained
from Eq. (\ref{GL CPS final}). If $\bB=B(0,0,1)$, then
\begin{equation}
\label{Hc2 z}
    H_{c2,z}(T)=\frac{\hbar c}{2e}\frac{\alpha}{K_1}(T_c-T).
\end{equation}
If $\bB=B(\cos\varphi,\sin\varphi,0)$, we choose the gauge
$\bA=Bz(\sin\varphi,-\cos\varphi,0)$. The lowest eigenvalue of the
GL operator corresponds to the order parameter with no modulation
along the field direction:
$$
    \eta(\br)=\exp\left[i\frac{2e}{\hbar
    c}(\bB\times\br)_zz_0\right]f(z),
$$
where $z_0$ is an arbitrary parameter. The function $f(z)$
satisfies an equation which can be reduced to the standard
harmonic oscillator equation by a shift in the coordinate: $z=
Z+z_0+(\hbar c\tilde K/4eK_1)$. Thus we find
\begin{equation}
\label{f z}
    f(z)\propto \exp\left(-\frac{eB}{\hbar
    c}\sqrt{\frac{K_1}{K_2}}\;Z^2\right),
\end{equation}
and the field-dependent critical temperature
\begin{equation}
\label{TcB CPS}
    T_c(\bB)=T_c(\bB=0)-\frac{2e}{\hbar
    c}\frac{\sqrt{K_1K_2}}{\alpha}B+\frac{\tilde K^2}{4\alpha K_1}B^2,
\end{equation}
which is completely isotropic in the $xy$-plane. We see that,
surprisingly, the $\tilde K$-term does not affect the linear in
$\bB$ suppression of $T_c$, giving rise only to a small, quadratic
in field, correction to $T_c(\bB)$. Neglecting the latter effect,
we find
\begin{equation}
\label{Hc2 xy}
    H_{c2,xy}(T)=\frac{\hbar c}{2e}\frac{\alpha}{\sqrt{K_1 K_2}}(T_c-T).
\end{equation}
The last term in Eq. (\ref{TcB CPS}) could become dominant in a
film of CePt$_3$Si. If the thickness of the film is less than the
correlation length $\xi_z=K_2/\alpha(T_c-T)$, then the order
parameter (\ref{f z}) becomes $z$-independent and the linear in
$\bB$ term in Eq. (\ref{TcB CPS}) is absent. Thus, in this case
the superconductivity can be promoted by a parallel magnetic
field, at least in the weak field limit. This agrees with the
results of Ref. \cite{Agter03}, where the gradient term linear in
$\bB$ and $\bm{D}$ was introduced on the phenomenological grounds
for a surface superconductor. The order parameter which occurs at
$T_c$ at non-zero $\bB$ is modulated in the $xy$ plane:
$\eta(\br)=\eta_0e^{i\bm{Q}\br}$, with $\bm{Q}\propto(\hat
z\times\bB)$ \cite{Agter03}, see also Ref. \cite{DF03}. It should
be noted though that the field-induced increase in $T_c$ may
indicate the onset of a magnetic instability of the
superconducting state, the investigation of which is beyond the
scope of the present work.

\section{Conclusions}
\label{sec: Conclusion}

We studied the magnetic properties of a clean superconductor with
spin-orbit coupling. We focussed on the weak-field limit near the
critical temperature, where the Ginzburg-Landau theory is
applicable. Starting from the effective single-band Hamiltonian in
the magnetic field, we obtained the expressions for the GL
effective masses and the internal magnetic moments of the Cooper
pairs in terms of the Fermi-surface averages, for an arbitrary
pairing symmetry and crystal structure, both in the
centrosymmetric and non-centrosymmetric cases.

For a superconductor without inversion symmetry, unusual terms,
linear in both the magnetic field and the order parameter
gradients, were found in the free energy expansion. The order
parameter itself corresponds to the pairing of electrons in the
time-reversed states within the same non-degenerate band. As a
simple application of our general formalism, we derived the GL
functional for CePt$_3$Si. It was found that although the unusual
gradient term does not affect the upper critical field in a bulk
sample, it could result in a field-induced enhancement of $T_c$ in
a thin film.

\section*{Acknowledgments}

The author is pleased to thank V. Mineev for the discussions which
initiated this project, D. Agterberg for valuable comments and
pointing out Refs. \cite{Agter03,DF03}, and B. Mitrovi\'c for
interest to this work. The financial support from the Natural
Sciences and Engineering Research Council of Canada is gratefully
acknowledged.

\appendix

\section{Derivation of Eq. (\ref{F r})}
\label{app: Proof F r}

To derive the free energy for a nonuniform distribution of the
order parameter, we start with a representation of the partition
function for the BCS Hamiltonian (\ref{H BCS 1}) in terms of a
functional integral over the Grassmann fields
$c_{\bk\alpha}(\tau)$ and $\bar c_{\bk\alpha}(\tau)$:
\begin{equation}
\label{Z gen}
    Z=\int{\cal D}c{\cal D}\bar c\;e^{-S},
\end{equation}
where $S=\int_0^\beta d\tau\left[\sum_{\bk}\bar
c_{\bk\alpha}\partial_\tau c_{\bk\alpha}+H(\tau)\right]$. The
interaction term in the action can be written as
$$
    S_{int}=-\frac{V}{4}\sum\limits_a\int_0^\beta d\tau\sum\limits_{\bq}
    B_a^\dagger(\bq,\tau)B_a(\bq,\tau),
$$
where
$$
    B_a(\bq,\tau)=\sum\limits_{\bk}\Psi_{a,\alpha\beta}^\dagger(\bk)
    c_{-\bk+\bq/2,\alpha}(\tau)c_{\bk+\bq/2,\beta}(\tau).
$$
The interaction term is then decoupled by means of the
Habbard-Stratonovich transformation, introducing a complex bosonic
field $\eta_a(\bq,\tau)$:
\begin{eqnarray*}
    e^{-S_{int}}\to\int{\cal D}\eta_a^*{\cal D}\eta_a\;
    \exp\Bigl\{-\sum\limits_a\int_0^\beta d\tau\sum\limits_{\bq}
    \Bigl[\frac{1}{V}|\eta_a|^2\\
    +\frac{1}{2}(B_a^\dagger\eta_a+\eta_a^*B_a)
    \Bigr]\Bigr\}.
\end{eqnarray*}
The last two terms in the exponent can be written as
\begin{eqnarray*}
    \frac{1}{2}\int_0^\beta d\tau\sum\limits_{\bk\bq}
    \Delta_{\alpha\beta}(\bk,\bq;\tau)\bar c_{\bk+\bq/2,\alpha}(\tau)
    \bar c_{-\bk+\bq/2,\beta}(\tau)\\+\mathrm{H.c.},
\end{eqnarray*}
where
\begin{equation}
\label{OP matrix}
    \Delta_{\alpha\beta}(\bk,\bq;\tau)=
    \sum\limits_a\eta_a(\bq,\tau)\Psi_{a,\alpha\beta}(\bk)
\end{equation}
is the order parameter matrix in the pseudospin space [cf. Eq.
(\ref{OP definition 1})].

The next step is to integrate out the fermionic degrees of
freedom, which can be achieved by using the four-component Nambu
spinor fields $C_{\bk}(\tau)=[c_{\bk\alpha}(\tau),\bar
c_{-\bk\alpha}(\tau)]^T$ and calculating a Gaussian fermionic
integral. As a result we arrive at the following representation of
the partition function:
\begin{equation}
\label{Z eta}
    Z=\int{\cal D}\eta_a^*{\cal
    D}\eta_a\;e^{-S_{eff}[\eta^*,\eta]},
\end{equation}
where
\begin{equation}
\label{S eff}
    S_{eff}=\frac{1}{V}\sum\limits_a\int_0^\beta d\tau\sum\limits_{\bq}
    |\eta_a|^2-\frac{1}{2}\mathrm{Tr}\ln(1-{\cal G}_0\Sigma)
\end{equation}
is the effective action for the superconducting order parameter.
Here ${\cal G}_0$ is the Gor'kov-Nambu Green's function at
$\eta=\eta^*=0$ (i.e. in the normal state):
\begin{equation}
\label{hat G0}
    {\cal G}_0=\left(%
\begin{array}{cc}
  G & 0 \\
  0 & -G^T \\
\end{array}%
\right),
\end{equation}
where $G=(-\partial_\tau-{\cal E})^{-1}$ is a $2\times 2$ matrix
in the pseudospin space, which satisfies Eq. (\ref{GFequation k}),
and $\Sigma$ is the $4\times 4$ matrix self-energy function
describing the superconducting pairing:
\begin{equation}
\label{Sigma}
    \Sigma=\left(%
\begin{array}{cc}
  0 & \Delta \\
  \Delta^\dagger & 0 \\
\end{array}%
\right),
\end{equation}
with the order parameter matrix defined by Eq. (\ref{OP matrix}).
The trace in the action (\ref{S eff}) should be understood as the
matrix trace in the four-dimensional Nambu $\times$ pseudospin
space, accompanied by the operator trace in the $\bk\tau$-space.

Using the partition function (\ref{Z eta}), we can calculate the
free energy of the system: ${\cal F}=-(1/\beta)\ln Z$. The BCS
mean-field approximation corresponds to a stationary saddle point
of the effective action (\ref{S eff}). For
$\eta_a(\bq,\tau)=\eta_a(\bq)$, the saddle-point action becomes
$S^{sp}_{eff}=\beta{\cal F}$, with the free energy (or, more
precisely, the difference between the free energies of the
superconducting and the normal states at the same temperature)
given by
\begin{equation}
\label{F gen}
    {\cal F}=\frac{1}{V}\sum\limits_a\sum\limits_{\bq}|\eta_a(\bq)|^2
    -\frac{1}{2\beta}\mathrm{Tr}\ln(1-{\cal G}_0\Sigma).
\end{equation}
The order parameter components satisfy the saddle-point equations
$\delta{\cal F}/\delta\eta_a^*=0$ (the GL equations). In the
vicinity of the critical temperature at arbitrary magnetic field,
the order parameter is small, so we can keep only the quadratic in
$\eta_a$ terms in the expansion of the trace in the free energy
(\ref{F gen}). In terms of the Fourier-transformed basis functions
\begin{equation}
    \Psi_{a,\alpha\beta}(\brho)=\sum\limits_{\bk}\,e^{i\bk\brho}
    \Psi_{a,\alpha\beta}(\bk)
\end{equation}
and the Green's functions (\ref{G rr def}), we have
\begin{equation}
\label{F}
    {\cal F}=\sum\limits_{ab}\int d\br_1d\br_2\;
    \eta_a^*(\br_1)S_{ab}(\br_1,\br_2)\eta_b(\br_2),
\end{equation}
with the kernel
\begin{widetext}
\begin{eqnarray}
\label{kernel gen}
 &&S_{ab}(\br_1,\br_2)=\frac{1}{V}\delta_{ij}\delta(\br_1-\br_2)\nonumber\\
 &&\qquad-\frac{1}{2}T\sum\limits_n\int d\brho_1 d\brho_2\;
 \Psi_{a,\alpha\beta}^\dagger(\brho_1)
 G_{\beta\gamma}\left(\br_1+\frac{\brho_1}{2},\br_2+\frac{\brho_2}{2};
 \omega_n\right) \Psi_{b,\gamma\delta}(\brho_2)
 G_{\alpha\delta}\left(\br_1-\frac{\brho_1}{2},\br_2-\frac{\brho_2}{2};
 -\omega_n\right).\qquad
\end{eqnarray}
\end{widetext}
Substitution of the factorized Green's function (\ref{GF
factorized 1}) in (\ref{kernel gen}) gives the phase factor
\begin{eqnarray*}
    &&\exp\Bigl[i\varphi\left(\br_1+\frac{\brho_1}{2},\br_2+
    \frac{\brho_2}{2}\right)+i\varphi\left(\br_1-\frac{\brho_1}{2},\br_2-
    \frac{\brho_2}{2}\right)\Bigr]\\
    &&=\exp\left[2i\varphi(\br_1,\br_2)+
    i\frac{e}{4\hbar c}\bB(\brho_1\times\brho_2)\right]
\end{eqnarray*}
[to prove this, one can use the Taylor expansions of the
$\varphi$'s with respect to $\brho_{1,2}$, and also the identities
(\ref{dvarphidr})]. The next step is to use
$$
    \exp\left\{i\frac{2e}{\hbar
    c}\int_{\br_1}^{\br_2}\bm{A}(\br)d\br\right\}\eta(\br_2)=
    e^{-i(\br_1-\br_2)\bm{D}_1}\eta(\br_1),
$$
where $\bm{D}=-i\nabla_{\br}+(2e/\hbar c)\bm{A}$, to cast the free
energy (\ref{F}) in the form (\ref{F r}), with the function $\bar
S(\bR)$ given by
\begin{widetext}
\begin{eqnarray}
\label{bar S R}
 \bar S_{ab}(\bR)=\frac{1}{2}T\sum\limits_n\int d\brho_1 d\brho_2\;
 \Psi_{a,\alpha\beta}^\dagger(\brho_1)\Psi_{b,\gamma\delta}(\brho_2)
 \exp\left[ i\frac{e}{4\hbar c}\bB(\brho_1\times\brho_2)\right]
 \nonumber\\
 \times \bar G_{\beta\gamma}\left(\bR+\frac{\brho_1-\brho_2}{2},\omega_n\right)
 \bar G_{\alpha\delta}\left(\bR-\frac{\brho_1-\brho_2}{2},-\omega_n\right).
\end{eqnarray}
\end{widetext}
Finally, taking the Fourier transform of this expression, we
arrive at Eq. (\ref{bar S}).

The analysis in the non-centrosymmetric case can be done in a
similar fashion, the only difference being that there is no
pseudospin degrees of freedom, and $G$, $\Psi$, and $\Delta$
become just scalar functions. The partition function still has the
form (\ref{Z eta}), but the effective action now reads
\begin{equation}
\label{S eff 2}
    S_{eff}=\frac{1}{2V}\sum\limits_a\int_0^\beta d\tau\sum\limits_{\bq}
    |\eta_a|^2-\frac{1}{2}\mathrm{Tr}\ln(1-{\cal G}_0\Sigma),
\end{equation}
where ${\cal G}_0$ and $\Sigma$ are $2\times 2$ matrix operators
in the Nambu space and the $\bk\tau$-space. Repeating all the
steps leading to Eq. (\ref{bar S R}), we arrive at Eqs. (\ref{S
2}) and (\ref{bar S 2}).

\section{Gradient energy near $T_c$}
\label{app: Linear B}

In this Appendix we estimate the lowest eigenvalue of the matrix
differential operator $\hat{\cal K}_{ab}=K_{ab,ij}D_iD_j$, where
$K_{ab,ij}$ are constant coefficients, $a,b=1..d$, and
$i,j=x,y,z$. We choose $\bB$ along the $z$ axis, i.e. $\bB=B\hat
z$ (one can always achieve that by rotating the coordinate system,
which is equivalent to a re-definition of $K_{ab,ij}$). It is
convenient to introduce new operators
\begin{equation}
\label{a operators}
    \left.\begin{array}{l}
    \displaystyle a_\pm=\frac{1}{2}\sqrt{\frac{\hbar c}{eB}}(D_x\pm iD_y),\\
    \\
    \displaystyle a_3=\sqrt{\frac{\hbar
    c}{eB}}D_z.
    \end{array}\right.
\end{equation}
It is easy to check that the operators $a_\pm$ satisfy the
relations $a_+=a_-^\dagger$ and $[a_-,a_+]=1$, and therefore have
the meaning of the lowering and the raising operators
respectively, while the operator $a_3=a_3^\dagger$ commutes with
both of them: $[a_3,a_\pm]=0$. Representing $\hat{\cal K}_{ab}$ in
terms of the operators (\ref{a operators}), we have
\begin{equation}
\label{cal K a}
    \hat{\cal K}_{ab}=\frac{eB}{\hbar c}\sum\limits_{n,m=\pm,3}\tilde
    K_{ab,nm}a_n^\dagger a_m,
\end{equation}
where the coefficients $\tilde K_{ab,nm}$ are linear combinations
of $K_{ab,ij}$ and therefore do not depend on $B$. It immediately
follows from the last expression that all eigenvalues of
$\hat{\cal K}$ are linear in $B$.

To calculate the eigenvalues explicitly, it is convenient to
choose the basis of states $|N,p\rangle$ such that
\begin{eqnarray*}
    &&a_+|N,p\rangle=\sqrt{N+1}|N+1,p\rangle\\
    &&a_-|N,p\rangle=\sqrt{N}|N-1,p\rangle\\
    &&a_3|N,p\rangle=p|N,p\rangle,
\end{eqnarray*}
where $N=0,1,...$ has the meaning of the Landau level index and
$p$ is a real number which is proportional to the wave vector
along the $z$-axis: $p=k_z\sqrt{\hbar c/eB}$. Expanding the
eigenfunctions of $\hat{\cal K}$ in this basis:
$\eta_a(\br)=\sum_{N,p}C_{a,N,p}\langle\br|N,p\rangle$, we arrive
at a system of linear equations for the coefficients $C_{a,N,p}$,
which is infinite in general. The upper critical field then
corresponds to the minimum eigenvalue of this system with respect
to $p$ (while it is usually assumed that the minimum is achieved
for $p=0$, some exceptions are discussed, e.g. in Ref.
\cite{LZh95}).

In some simple cases, the diagonalization procedure outlined above
can be carried out analytically. For example, for a one-component
order parameter in an isotropic $s$-wave superconductor we have
\begin{equation}
\label{cal K s wave}
    \hat{\cal K}=K(D_x^2+D_y^2+D_z^2)=\frac{eB}{\hbar c}
    K(4a_+a_-+a_3^2+2).
\end{equation}
Since $a_+a_-|N,p\rangle=N|N,p\rangle$, we have
\begin{equation}
    \hat{\cal K}|N,p\rangle=\frac{eB}{\hbar c}
    K(4N+p^2+2)|N,p\rangle.
\end{equation}
The lowest eigenvalue corresponds to $N=p=0$, which gives the
standard expression for the critical temperature suppressed by the
field:
\begin{equation}
    T_c(\bB)=T_c(\bB=\bm{0})-\frac{2e}{\hbar c}\frac{K}{\alpha}B.
\end{equation}

\end{document}